\documentclass[pra,letterpaper,twocolumn,showpacs,superscriptaddress,floatfix]{revtex4}
\usepackage{graphicx,psfrag,amsmath,amssymb,amsfonts,bbm,latexsym,color,dcolumn,bm}

\begin{document}

\title{Squeezing and robustness of frictionless cooling strategies}

\author{Stephen Choi}

\affiliation{Department of Physics, University of Massachusetts, Boston, MA 02125, USA}

\author{Roberto Onofrio}

\affiliation{Dipartimento di Fisica e Astronomia ``Galileo Galilei'', Universit\`a  di Padova, 
Via Marzolo 8, Padova 35131, Italy}

\affiliation{ITAMP, Harvard-Smithsonian Center for Astrophysics, 60 Garden Street, Cambridge, MA 02138, USA}

\author{Bala Sundaram}

\affiliation{Department of Physics, University of Massachusetts, Boston, MA 02125, USA}

\date{\today}

\begin{abstract}
Quantum control strategies that provide shortcuts to adiabaticity are increasingly 
considered in various contexts including atomic cooling. Recent studies have emphasized practical issues in order 
to reduce the gap between the idealized models and actual ongoing implementations. 
We rephrase here the cooling features in terms of a peculiar squeezing effect, and use it 
to parametrize the robustness of  frictionless cooling techniques with respect to noise-induced
deviations from the ideal time-dependent trajectory for the trapping frequency. We finally discuss 
qualitative issues for the experimental implementation of this scheme using bichromatic optical 
traps and lattices, which seem especially suitable for cooling Fermi-Bose mixtures and for investigating 
equilibration of negative temperature states, respectively.
\end{abstract}

\pacs{37.10.De, 42.50.-p,37.10.Vz,05.30.Fk}

\maketitle

\section{Introduction}
Quantum control techniques that provide shortcuts to adiabaticity, 
recently proposed in the atomic physics framework \cite{Chen}, are experiencing rapid growth due to the necessity, in various contexts, to reduce the 
undesired effects of a prolonged exposure to the external environment. Such technique produces the same {\it net} result as if a full adiabatic 
following has been implemented, but on a much shorter timescale.  More specifically, in a shortcut to adiabaticity technique
the composition of eigenstates one started with is preserved at the end of the process, but not 
{\it during} the process, unlike full adiabatic following. The first experimental 
implementations have demonstrated efficient fast decompression of ${}^{87}$Rb atoms in 
normal \cite{Schaff0} and Bose-condensed \cite{Schaff} states, as well as fast atomic 
transport \cite{Torrontegui}. 
Consequently, a number of contributions have focused attention on possible deviations from 
ideal realizations which could constitute bottlenecks to further implementations of these 
techniques.  In one study, it was found that the transient energy excitation limits the maximum attainable 
speed \cite{Chen1}. In another, deviations from harmonicity in the trapping potential, with 
specific attention to the crosstalk between the dynamics in different trapping directions, 
have  recently been discussed \cite{Torrontegui1}. 

Our previous work \cite{Choi} showed that frictionless cooling (also known as ``fast adiabatic" cooling), an example of shortcut to adiabaticity technique, may be optimal 
in controlling the temperature of a  buffer gas in the sympathetic cooling of another species, such 
as a Fermi gas. This is because this cooling method allows for retaining the maximum value of 
the gas heat capacity, both as the gas does not enter the degenerate regime, and as it  avoids the 
intentional loss of atoms required in evaporative cooling.
Even in this case, we found that practical issues may limit the successful application of this technique. 
Specifically, it was shown that, for short cooling times, atoms temporarily attain 
temperatures higher than the initial temperature before settling down to the final temperature. 
This can potentially result in atom losses in any realistic potential with a finite trapping depth. 
Also, very small final temperatures may result in a large spatial mismatch between the sizes of 
the two clouds making sympathetic cooling less effective. These considerations put practical restrictions 
on this method. 

Here, we extend our analysis by studying the robustness of the 
frictionless cooling method with respect to the likely presence of noise in a potential experiment. 
This is in light of the fact that, experimentally, there may be systematic sources of error which 
are of a truly {\sl nonlinear} nature, such as noise in the trapping frequency or 
anharmonicities in the trapping potential. To model the noise, we assume that an experimentalist has 
a well-defined protocol for implementing frictionless cooling and that the effect of various noise 
sources present in the apparatus (such as current noise in the coils for magnetic trapping, or 
laser power fluctuations in an optical dipole trap) add up to deviations from the ideal (Ermakov) trajectory.

In order to parametrize the robustness, one needs to estimate how much the eigenstate composition of the final state 
deviates from that of the initial state. 
However,  usual  measures such as the fidelity of the wave function 
${\cal F} = |\langle \psi (t_{\mathrm f})|\psi(0)\rangle|$  are unsuitable  in this scheme since, due 
to the trap relaxation, the final wave function spreads out in space significantly.  
To circumvent this constraint, we note that our particular frictionless cooling 
method\cite{Choi} can be formally linked to squeezing, and this allows us to use the amount of 
squeezing produced as a measure of  robustness. In particular, starting from a minimum uncertainty state, 
frictionless cooling should result in a minimum uncertainty state at the end of the run. 
Any deviation may be considered an indicator of less than perfect recovery of the 
desired final state, {\it i.e.}  a sign of an imperfect ``adiabatic following.''  
It should be noted that there are limitations 
in using squeezing as a general measure of robustness in situations other than the 
specific cooling scheme under consideration. Specifically, one cannot 
easily generalize this measure to more sophisticated states, such as mixed states 
or non-minimal uncertainty states. As a supplementary figure of merit of robustness, we  have also chosen a modified
fidelity measure ${\cal F'} = |\langle \psi(t_f) | \psi_{target} \rangle |$ where  $\psi(t_f)$  and    $\psi_{target}$ are the final states attained in the presence  and in the absence of noise, respectively, and present 
results using this measure along with the squeezing parameter.

The paper is organized as follows: In Section II we discuss the connection of frictionless
cooling to squeezing which  we use as a useful indicator of the fidelity in reaching a targeted final 
temperature. In Section III we apply these considerations to discuss the robustness of 
the ideal {shortcut to adiabaticity} trajectories to noise fluctuations in the trapping parameters. 
The possible noise sources are characterized as broadband and monochromatic, and we 
discuss the interplay between their spectra in relation to the intrinsic spectral content of the 
 frictionless cooling protocol. We finally provide various connections to
recent developments in the control of ultracold atoms, including a
qualitative discussion of the realization of  frictionless cooling in
a bichromatic optical trap, a promising candidate 
for implementing the antitrapping stage required for very  fast cooling, provided that 
noise fluctuations in the trap are kept under control.

\section{Squeezing in  frictionless cooling}

As discussed in Ref. \cite{Chen}, frictionless cooling relies on  specific dynamics 
of the harmonic trapping frequency to transfer  a quantum state such as an equilibrium thermal state
in a trap of initial trapping  angular frequency $\omega_0$ to that of final  angular 
frequency $\omega_{\mathrm f} \ll \omega_0$. This ensures that the energy level spacings in 
the final trap are much smaller than those of the initial trap, hence lowering the overall temperature. 
As we have discussed in detail in our previous work \cite{Choi}, we have additionally used 
the momentum variance as an indicator of temperature and have shown that the final wave function 
has its momentum variance reduced by a factor of $\omega_0/\omega_{\mathrm f}$ compared to that of 
the initial wave function.  The use of the momentum variance as an indicator of the temperature
in an ultracold atomic cloud has been recently validated in dedicated experiments \cite{Davis}.

The shortcut to adiabaticity is  accomplished using a method originally studied by Lewis and 
Riesenfeld \cite{Lewis0,Lewis}, who introduced the invariant operator 
related to the harmonic oscillator with an arbitrary time-dependent harmonic potential, 
$\hat{I}(t)=(b\hat{p}-m \dot{b}\hat{q})^2/2m+ m \omega_0^2 \hat{q}^2/(2 b^2)$. A key feature  is that the
eigenstates of $\hat{I}(t)$ can be made to coincide with the eigenstates of the harmonic oscillator 
Hamiltonian, particularly at the beginning  and at the end of the cooling process.  
Here  $b(t)$ is a time-dependent frequency scaling factor which satisfies the Ermakov equation 
$\ddot{b}(t) + \omega(t)^2b = \omega_0^2/b^3$. This can be solved by imposing boundary conditions 
on $b(t)$ and its first and second time derivatives, while assuming both a targeted final trapping 
frequency $\omega_{\mathrm f}$, and a total time duration for the protocol, $t_{\mathrm f}$ \cite{Chen}. 
This results in  a well-defined Ermakov trajectory for the trap frequency  
$\omega(t)$. By constructing a harmonic oscillator with the Ermakov trajectory for 
its time-dependent trapping frequency,  a ground state experiences dynamic changes 
in the trapping frequency that eventually results in its cooling within the specified 
time $t_{\mathrm f}$.  In this paper, we shall denote  the Ermakov trajectory as $\omega_{E}(t)$  
in order to distinguish it from a general time-dependent trap frequency.

Formally, the  above method  may be understood as a transformation of a harmonic oscillator 
Hamiltonian to the space of the invariant operator $\hat{I}(t)$. In particular, the net 
effect can be considered as an added nonlinearity to the Hamiltonian that produces
squeezing of atomic states. A first step in
this direction has been reported in \cite{Sokolovski} where number (Fock) states during sudden and
adiabatic expansions of the trap have been discussed. One can show
how squeezing is a natural byproduct of
the Ermakov dynamics by noting the relationship between the raising 
and lowering operators of the harmonic oscillator Hamiltonian  $a^{\dagger}(t)$ and $a(t)$, and
the Dirac raising and lowering operators for the Lewis and Riesenfeld invariant  operator $\hat{I}(t)$ 
 $\alpha^{\dagger}(t)$  and $\alpha(t)$\cite{Lewis}:
\begin{eqnarray}
\alpha(t)    &= &  \eta(t)  a(t)  + \zeta(t)  a^{\dagger}(t) 
\label{Bog1} \\ 
\alpha^{\dagger}(t)  & = &    \zeta(t)^{*}  a(t)   + \eta(t)^{*}  a^{\dagger}(t)    
\label{Bog2}
\end{eqnarray}
where
\begin{eqnarray}
\eta(t)  & = &   \frac{1}{2}  \left [ \sqrt{\frac{ \omega_0}{\omega_{\mathrm f}}} \frac{1}{b(t)}  
+  \sqrt {\frac{\omega_{\mathrm f}}{ \omega_0}}  b(t) - i  \frac{ \dot{b}(t)  }
{\sqrt{\omega_{\mathrm f} \omega_0}}   \right ]  ,  \\
\zeta(t)  & = &    \frac{1}{2} \left [ \sqrt{\frac{ \omega_0}{\omega_{\mathrm f}}} \frac{1}{b(t)}  
- \sqrt{\frac{ \omega_{\mathrm f}}{\omega_0}}  b(t)  - i \frac{\dot{b} (t) }{\sqrt{\omega_{\mathrm f} 
\omega_0}} \right ].  
\label{eta} 
 \end{eqnarray}
 
Despite the complicated expressions for $\eta(t)$ and $\zeta(t)$,  Eqs. (\ref{Bog1}) and  (\ref{Bog2}) are 
essentially Bogoliubov transformations, well-known in the description of squeezing in quantum optics \cite{QO}. 
In particular, $\eta(t)$ and $\zeta(t)$ satisfy  the condition $|\eta|^2 - |\zeta|^2 =  1$, 
{\it i.e.}   one can  directly recover  the standard squeezed state result of quantum optics 
\cite{QO} and write $\alpha = S a  S^{\dagger}$ where  $S$ is a unitary squeezing operator 
$S(\xi) = \exp \left ( -\xi a^{\dagger 2}/2 +  \xi^{*} a ^{2}/2  \right )$ with $\xi = 
\cosh^{-1} |\eta(t)|$ which, acting on a vacuum, produces a squeezed state  
$| \xi \rangle   =  S(\xi) | 0 \rangle$.  This connection went 
unnoticed in the paper by Lewis and Riesenfeld \cite{Lewis}, presumably because the field of quantum optics 
was still in its infancy.
 
While  the position variance of the wave function undergoing the  frictionless cooling is analytically 
given as a  function of the solution to the 
Ermakov equation $b(t)$ \cite{Chen}  
\begin{equation}
\sigma^2_{x}(t)  = \frac{\hbar}{2 m \omega_0} b^{2}(t),   \label{pos_var}
\end{equation}
it can be shown that the momentum variance, proportional to the temperature of the gas,  evolves as 
\begin{equation}
\sigma^2_{p}(t)  = \frac{\hbar m \omega_{\mathrm f} }{2}  \exp  (2 \cosh^{-1} |\eta(t)| ) ,
\end{equation}
so that their product is
\begin{equation}
\sigma^2_{x}(t) \sigma^2_{p}(t)  = \frac{\hbar^2}{4} \frac{\omega_{\mathrm f}}{\omega_0}  b^2(t)
\exp   ( 2 \cosh^{-1} |\eta(t)|  ) .   
\label{varprod}
\end{equation}
This shows that  the frictionless cooling
strategy formally produces a squeezed state, and may be potentially useful as a quantum control method 
for generating squeezing in cold atoms.  At the final time $t = t_{\mathrm f}$, 
$b(t_{\mathrm f}) =  \sqrt{\omega_0/\omega_{\mathrm f}}$ and $\exp(2 \cosh^{-1} |\eta(t_{\mathrm f})|) = 1$, 
such that  $\sigma^2_{x}(t_{\mathrm f}) \sigma^2_{p}(t_{\mathrm f}) =  \hbar^2/4$ 
{\it i.e.}  the state returns to the minimum  uncertainty state as desired. 
This implies that one may quantify the robustness to external noise by checking 
how much the state differs from the minimum uncertainty state when the time-dependent 
trap frequency deviates from the optimal Ermakov trajectory.  

To investigate quantitatively the squeezing effect in frictionless cooling strategies, we have numerically 
solved the Schr\"{o}dinger equation for the time-dependent harmonic oscillator with the trap 
frequency given by $\omega_{E}(t)$. As a starting point, we have compared the two decompression 
strategies for a harmonically trapped atomic cloud we have used in our previous work \cite{Choi} 
where, for concreteness, we assumed the initial trap  angular frequency $\omega_0 = 2 \pi \times 250$ Hz and  
the final  trap angular frequency $\omega_{\mathrm f} = 2\pi  \times 2.5$ Hz. 
The two strategies are characterized by their duration:  $ t_{\mathrm f}$=25 ms for which  
$\omega^2_{E}(t) > 0$ throughout the entire run, and 
$t_{\mathrm f}$=6 ms which includes a short time interval for which $\omega^2_{E}(t)<0$.
During this  interval  the atoms temporarily experience an {\sl antitrapping} potential 
 (like in an inverted harmonic oscillator, with the possibility of cooling first 
noticed in the conclusions of \cite{Yuce}) leading to an accelerated spreading of 
the wave function, after which the trap is flipped back to allow regrouping, and 
desired cooling, of the wave function by $t = t_{\mathrm f}$.

\begin{figure}[t]
\begin{center}
\includegraphics[width=0.9\columnwidth]{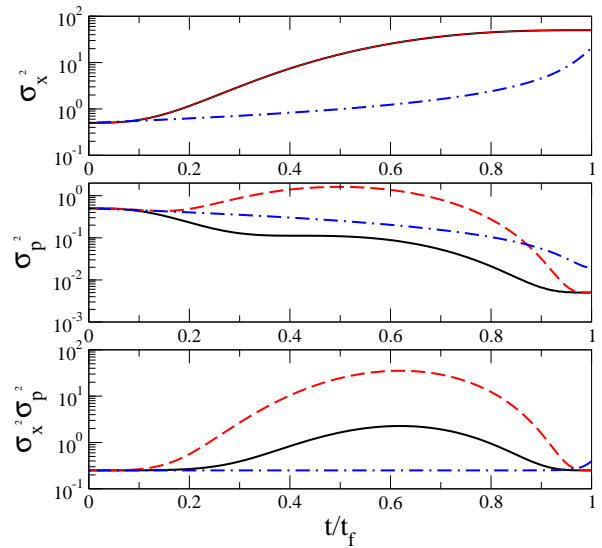}
\end{center}
\caption{(Color online) Squeezing properties of the frictionless cooling strategies. 
Depicted is the time dependence of the position variance (top), 
momentum variance (center), and their product (bottom) for the three 
cooling strategies discussed in the text, with the product $\sigma_x^2 \sigma_p^2$ 
expressed, from now on, in units of $\hbar^2$. The dashed and solid lines are for 
 frictionless cooling with $t_{\mathrm f}$=6 ms and 25 ms, respectively. 
The dot-dashed line is for a linear ramp-down of the frequency
corresponding to a quasi-adiabatic expansion occurring in a time
$t_{\mathrm f}=2 \pi/\omega_{\mathrm f}$ as discussed in \cite{Choi}. 
The adiabatic invariant strategies allow for the preservation of
states with minimal uncertainty product, as they display an uncertainty product
returning to the value $\sigma_x^2 \sigma_p^2=\hbar^2/4$ at
$t=t_{\mathrm f}$ after a transient increase at earlier times. 
Conversely, the quasi-adiabatic expansion keeps its uncertainty
product constant for a long time interval, but manifests an
increase approaching the final time as the quasi-adiabaticity 
condition is less satisfied in the latest stage with minimal trapping frequency.}
\label{adiabaticfig1}
\end{figure}

We show in  Fig. 1 the time dependence of the position and 
momentum variances as well as their product for the two durations of cooling. 
The temporal variation of the position variance is found to be  independent of $t_{\mathrm f}$, as expected 
from the  analytical results, while the time evolution of the product of variances  clearly indicates 
the recovery of the minimum uncertainty state at $t = t_{\mathrm f}$. For comparison, a quasi-adiabatic 
trajectory obtained by a linear ramping down of the frequency to its final value is also depicted, 
showing that the uncertainty product in this case exceeds the minimum value at final time.

\begin{figure}[t]
\begin{center}
\includegraphics[width=0.4\textwidth]{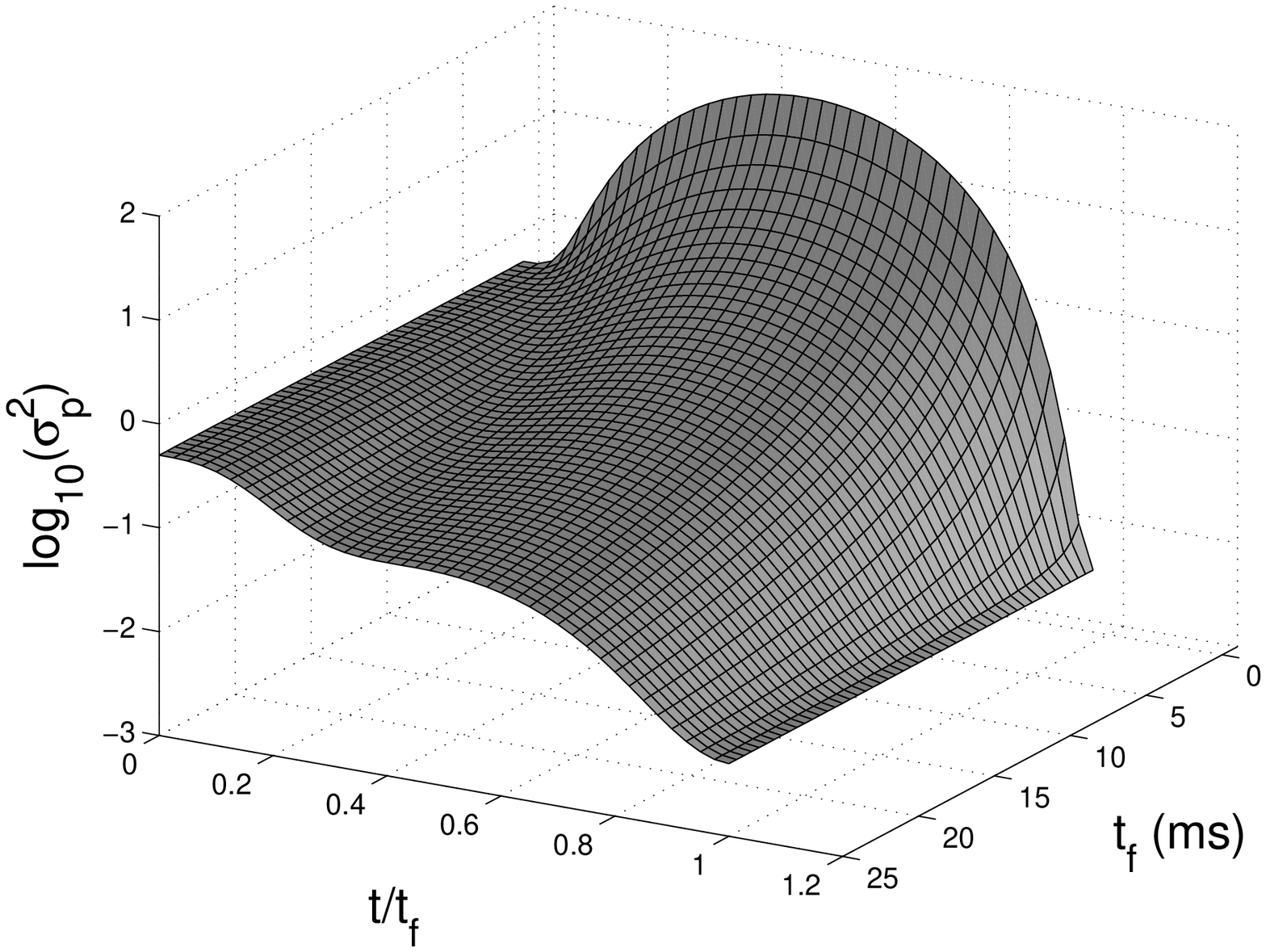} 

\includegraphics[width=0.4\textwidth]{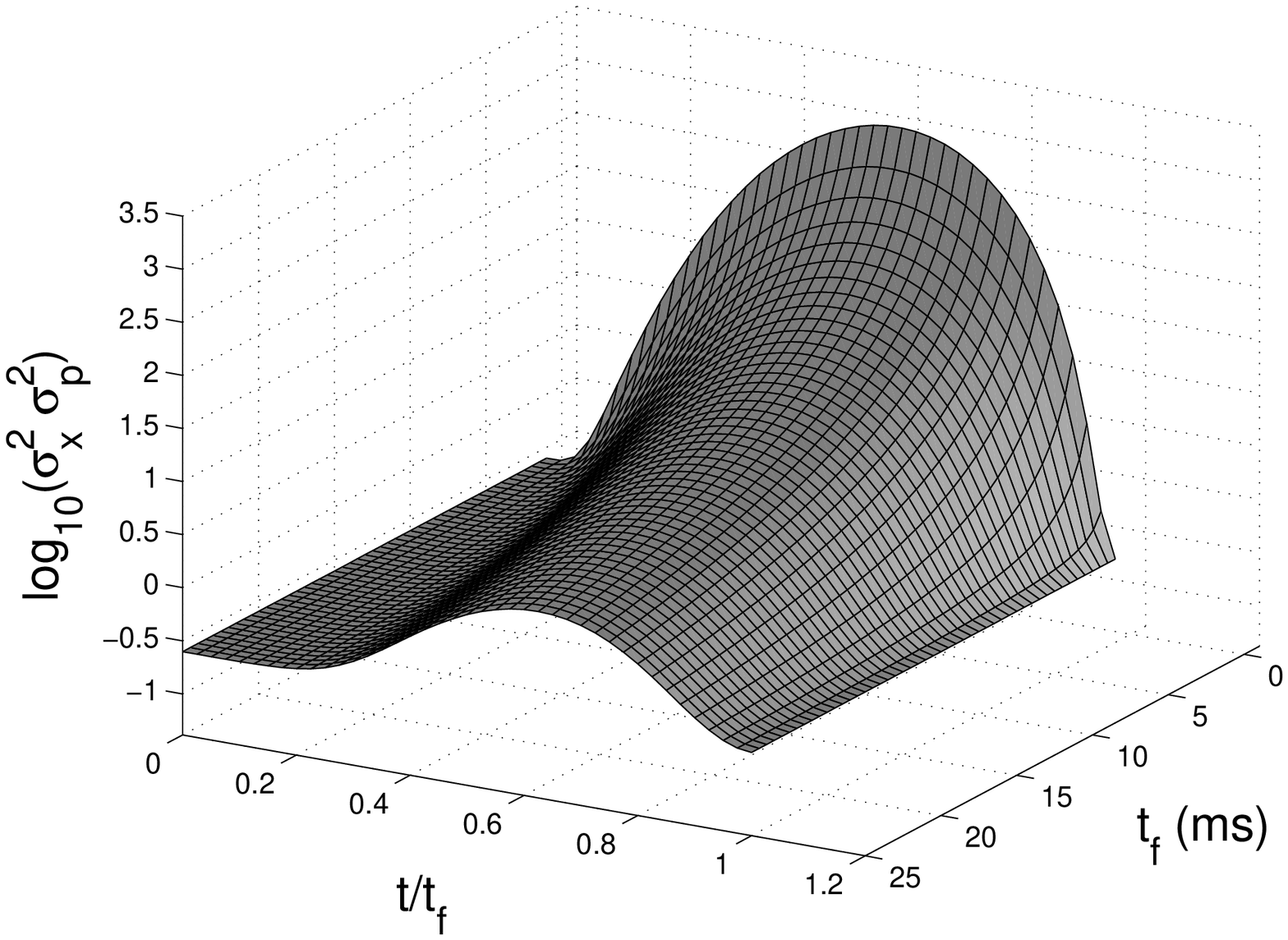} 

\end{center}
\caption{Top: $\log_{10} \sigma^2_{p} (t) $ calculated directly from the 
squeezing transformation for various $t_{\mathrm f}$. Bottom: $\log_{10} [ \sigma^2_{x} (t) \sigma^2_{p} (t)]$   
 for various $t_{\mathrm f}$ indicating the deviation from, and the eventual 
return to, the minimum uncertainty state over time. It is noted that for small $t_{\mathrm f}$, 
the momentum variance $\sigma^2_{p} (t)$ proportional to the temperature, goes up significantly 
during the run. As  previously noticed in our work \cite{Choi}, due to potential atom loss, this presents 
a restriction on using this method in sympathetic cooling. A consequence relevant to this paper 
is the very high values of the product $\sigma^2_{x} (t) \sigma^2_{p} (t)$ during the run for short $t_{\mathrm f}$.}
\label{adiabaticfig2}
\end{figure}

It should be noted here that, since Eqs.  (\ref{pos_var})-(\ref{varprod}) are explicitly given in terms of the 
Ermakov solution $b(t)$, it is possible to directly calculate the amount of squeezing without evolving 
the wave function using the Schr\"{o}dinger equation with a time-dependent potential. 
Figure 2 shows the momentum  variance and the product of position and momentum variances 
as a function of time for various possible values of $t_{\mathrm f}$.  
The results for  $t_{\mathrm f} $=6 ms and  $t_{\mathrm f} $=25 ms 
obtained by evolving the Schr\"{o}dinger equation, were found accurately reproduced. 
It is clear that the methods of quantum optical squeezing are applicable in the context 
of  frictionless cooling, although the system is rather different from the standard quantum 
optical systems involving nonlinear optical media.  

For  small $t_{\mathrm f}$, it is noted that the momentum variance $\sigma^2_{p} (t)$ 
goes up significantly during the run, resulting in large values of the variance product, 
$\sigma^2_{x} (t) \sigma^2_{p} (t)$.   The presence of such large deviation from the minimum 
uncertainty state during the run implies that, for such cases, any noise is likely to  more 
effectively push the system away from the minimum uncertainty state.  
Indeed, we observe such a trend in the next section.

\begin{figure}[t]
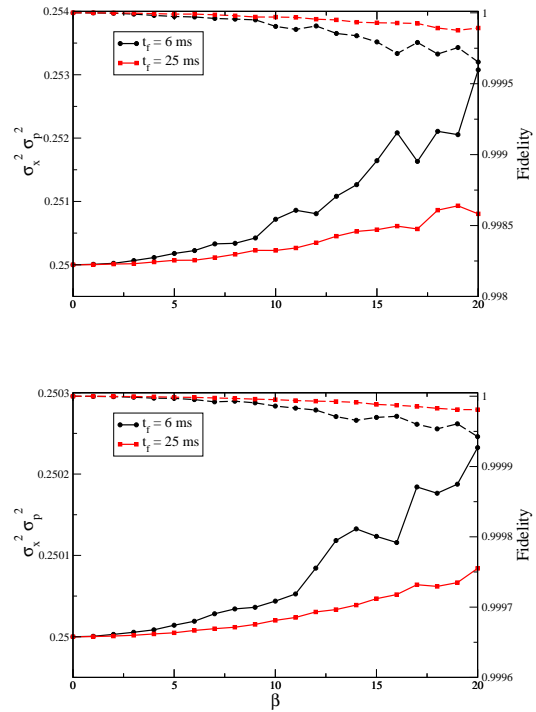

\begin{center}
\includegraphics[width=0.8\columnwidth]{adiabaticnewfig3a.eps}

\vspace{1cm}

\includegraphics[width=0.8\columnwidth]{adiabaticnewfig3b.eps}
\end{center}
\caption{(Color online) The final variance product, 
$\sigma^2_{x} (t_{\mathrm f} ) \sigma^2_{p} (t_{\mathrm f})$ (bottom solid line, left vertical axis)
 and the fidelity (top dashed line, right vertical axis) for $t_{\mathrm f}$=25 ms (red squares) and $t_{\mathrm f}$=6 ms 
(black circles)  as a function of noise amplitude $\beta$ described in the text. Gaussian white noise  (top plot); 
uniform white noise (bottom plot), all from the averaging of 150 realizations.}
\label{adiabaticfig3}
\end{figure}

\begin{figure}[t]
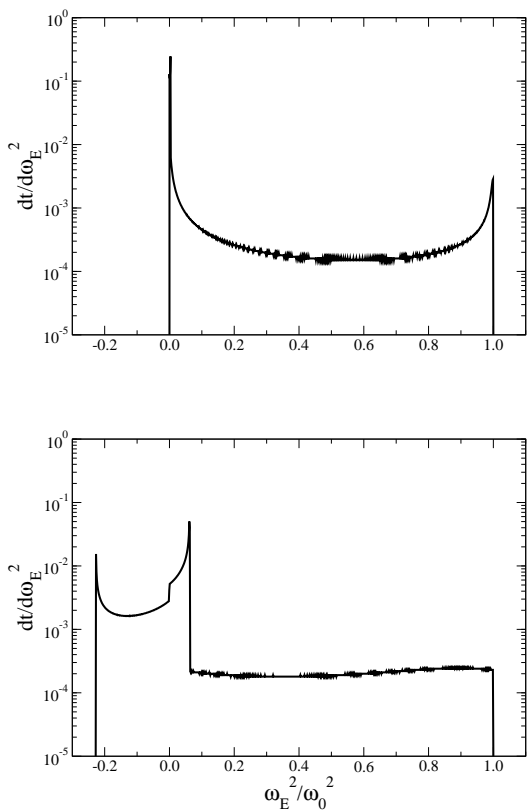

\begin{center}
\includegraphics[width=0.8\columnwidth]{adiabaticnewfig4a.eps}

\vspace{1cm}

\includegraphics[width=0.8\columnwidth]{adiabaticnewfig4b.eps}
\end{center}
\caption{Temporal density of states versus $\omega^2/\omega_0^2$ for
the two adiabatic strategies discussed in the text with $t_{\mathrm f}$=25 ms (top) 
and 6 ms (bottom). The temporal density of states,
here evaluated for $10^5$ samples and $10^3$ squared frequency
intervals in the entire range of variability of $\omega^2$, is 
normalized to unity, thereby representing the probability density 
to realize a given $\omega^2$ during the corresponding frictionless cooling process.}
\label{adiabaticfig4}
\end{figure}

\section{Robustness of the frictionless cooling scheme}

In this section, we want to answer the following question: how robust is the frictionless cooling technique
to small deviations from the optimized Ermakov trajectory? Since, as discussed in the last section, 
the variable frequency strategy is basically mappable to a nonlinear system, the interplay between 
the deterministic Ermakov trajectory and any added noise may,  in principle, be subtle and can lead 
to large and detrimental deviations from the idealized behavior. To study robustness in this system, 
we consider two experimentally relevant types of noise, broadband and monochromatic.   
These choices allow us to understand the system  via the complementary responses.

\begin{figure*}[t]
\includegraphics[width=0.49\textwidth]{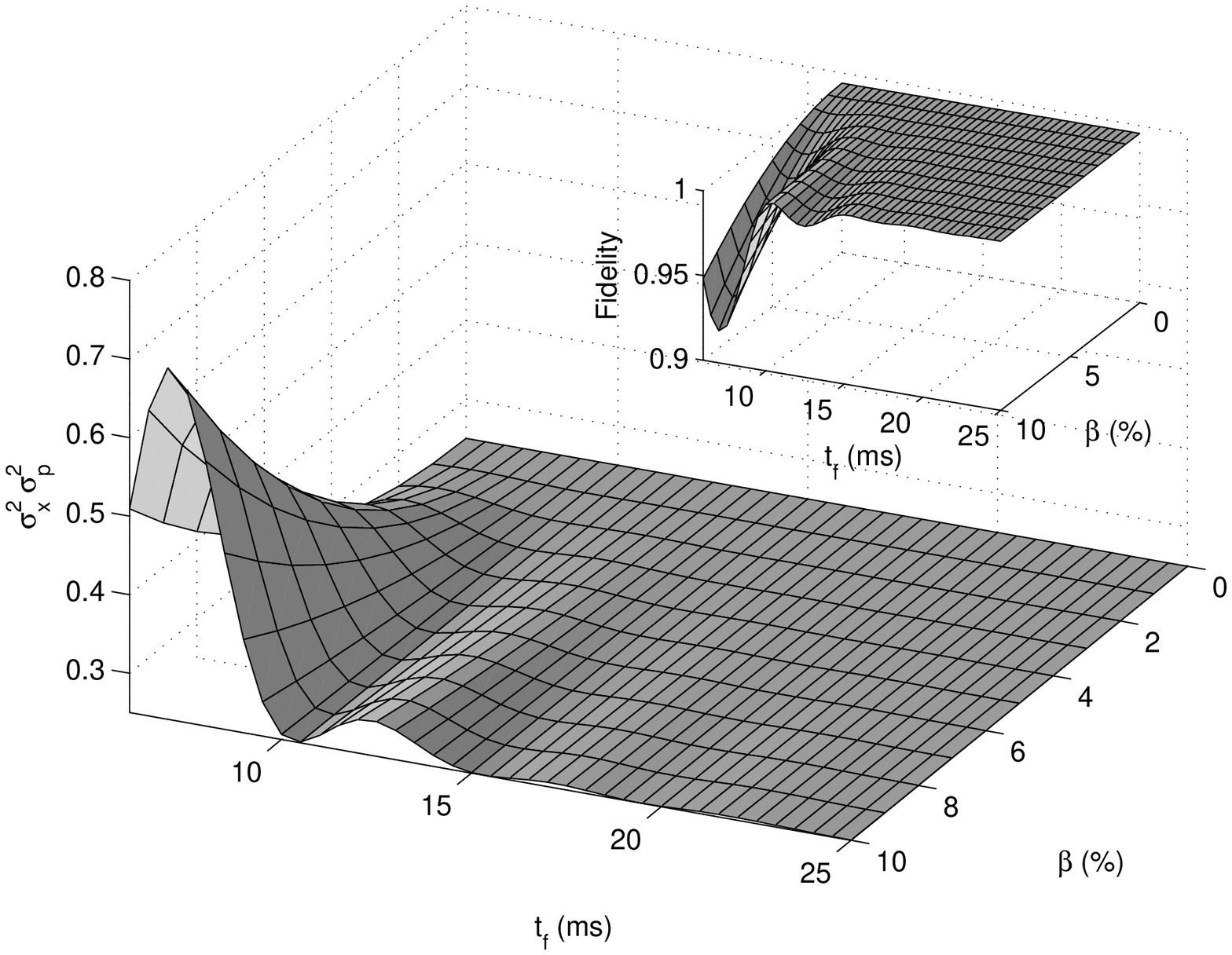}
\includegraphics[width=0.49\textwidth]{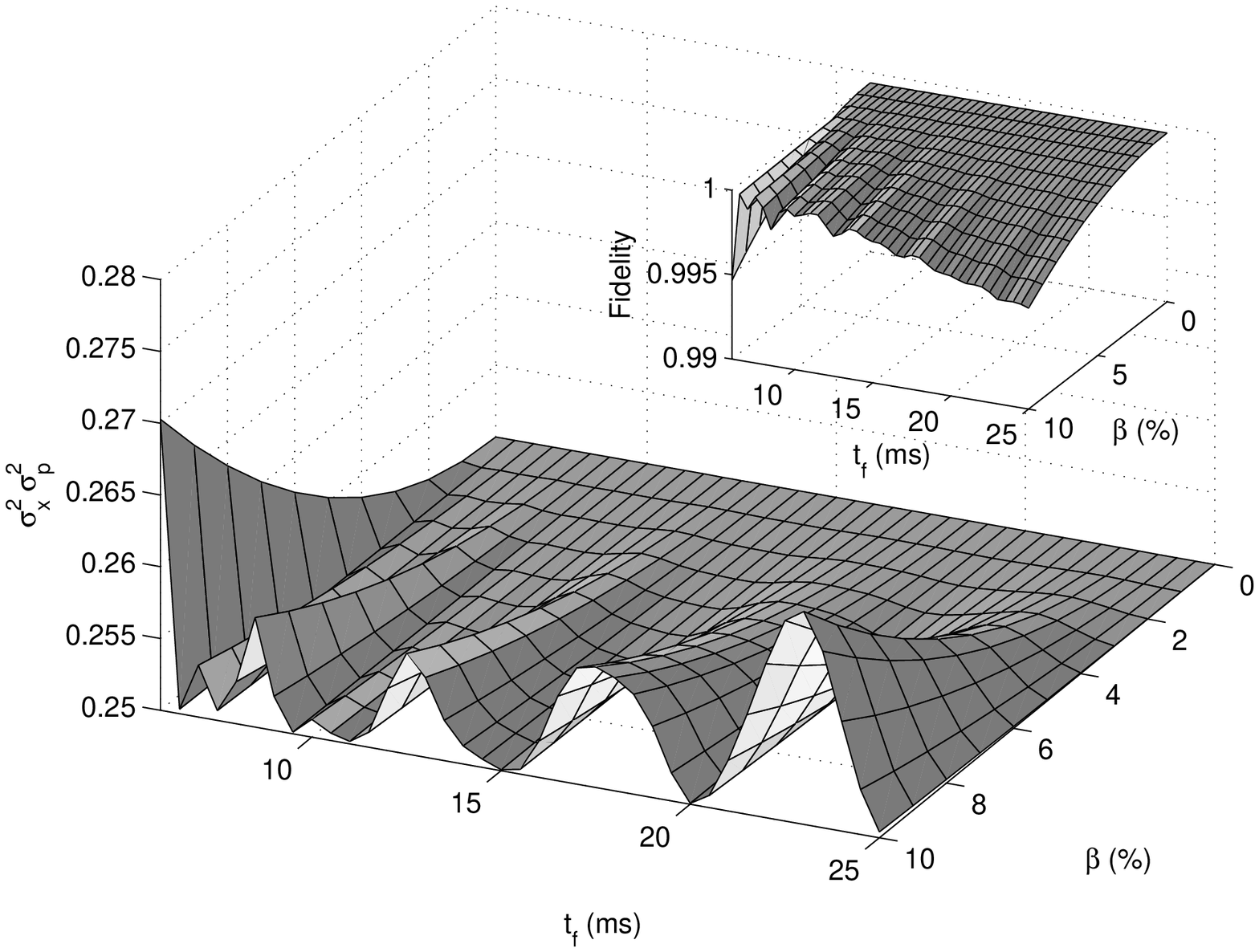}


\caption{The final variance product,  $\sigma^2_{x}(t_{\mathrm f} )
\sigma^2_{p}(t_{\mathrm f} )$   and the fidelity (inset) as a function of $t_{\mathrm f}$ and
the amplitude of sinusoidal modulation $\beta$ in a modulated
trajectory $\omega^2_{s}(t) =\omega^2_{E}(t) [1 + \beta
\sin(\Omega_{n} t)]$.  In the left panel we show the case of a 
sinusoidal noise within the range of trapping frequencies, 
$\Omega_{n}  = \omega_0/2$, in the right panel  an example of 
noise outside, $\Omega_{n}  = 3\omega_0/2$.}
\label{adiabaticfig5}
\end{figure*}

\subsection{Response to random noise}

For the random noise, we consider two different possibilities: Gaussian white noise with mean 
value $\omega^2_{E}(t)$, and standard deviation $\beta | \omega^2_{E}(t) |$, and uniformly 
distributed noise centered at $\omega^2_{E}(t)$ with width  $\beta | \omega^2_{E}(t) |$, where $\beta$ 
is varied from zero to  0.2, {\it i.e.} in both situations noise has an amplitude up to 20\% of 
the unperturbed trajectory $| \omega^2_{E}(t) |$.  The results of simulations with the Gaussian white 
noise and uniform white noise are shown in Fig. \ref{adiabaticfig3} where the final 
variance product,  $\sigma^2_{x}(t_{\mathrm f} ) \sigma^2_{p} (t_{\mathrm f} )$  and the fidelity, 
obtained by averaging 150 trajectories, is depicted.  Both the variance product and 
the fidelity show that the overall deviation  from the desired result is of order $0.1\%$, and hence the system is relatively insensitive to random noise.
This scheme is therefore more robust than we expected, especially given the very large 
amount of noise (up to 20\%) added. This may be a result of the versatility of the Ermakov
invariant scheme which, in principle, is able to handle all values of $t_{\mathrm f}$; the net effect 
of added noise may be viewed as shifting the final $t_{\mathrm f}$ which still produces the correct state.
As expected, the final state deviates more and more from the  desired state as one increases 
the noise amplitude $\beta$. It is worth noting that, as we had expected from examining Fig. 
\ref{adiabaticfig2}, the $t_{\mathrm f}=$6 ms case with negative square frequency in the 
Ermakov trajectory is found to be more sensitive to the effect of noise. Physically, this is 
reasonable considering the added vulnerability to noise during the time interval in which atoms 
experience their antitrapping stage.

\subsection{Response to sinusoidal modulation}

Noting that  realistic random noise can be viewed as a sum over sinusoidal modulations with 
random frequencies and amplitudes, one can analyze the robustness of this system by studying 
the response to a range of purely sinusoidal modulations. Sinusoidal modulations may occur as 
an unintended noise or added  intentionally, for instance to improve the signal-to-noise ratio 
of a specific feature under investigation, to create arbitrary trapping potentials \cite{Courteille}, 
or to prepare well-defined target states, for instance atomic Fock states in optical traps
\cite{Dudarev,Campo,Pons,Wan} and optical lattices \cite{Meisner,Nikolopoulos}.  
Continuing our analogy to quantum optics, the study of monochromatic noise is similar 
to examining the response of a nonlinear optical medium to coherent  light of fixed frequency.  

Before such analysis we need to specify the range of trapping frequencies involved in the 
cooling procedure. We indeed expect that the effect of noise sources will depend on their 
relative location with respect to the spectrum of frequencies of the system. To quantify the 
latter, we introduce a  `temporal' density of states defined as the amount of time spent by the system 
in the range between $\omega^2_E$ and $\omega^2_E+d(\omega^2_E)$. This is directly related, at 
least for monotonic decreases of the trapping frequency, to $[d(\omega^2_E)/dt]^{-1}$, the 
inverse of the slope of the curve $\omega^2_{E}(t)$. For the case of nonmonotonic behavior, 
such as in the strategies involving a stage in which $\omega^2_E <0$, this is less trivial 
to obtain analytically. However, in practice, we can use a simple code in which we
sort the times at which frequencies occur and count the number of times we find 
a square frequency in the $[\omega^2_E, \omega^2_E+d(\omega^2_E)]$ interval. For the 
 frictionless cooling strategy with $t_{\mathrm f}$=25 ms this is depicted in the top 
plot of Fig. 4. 
\begin{figure}[t]
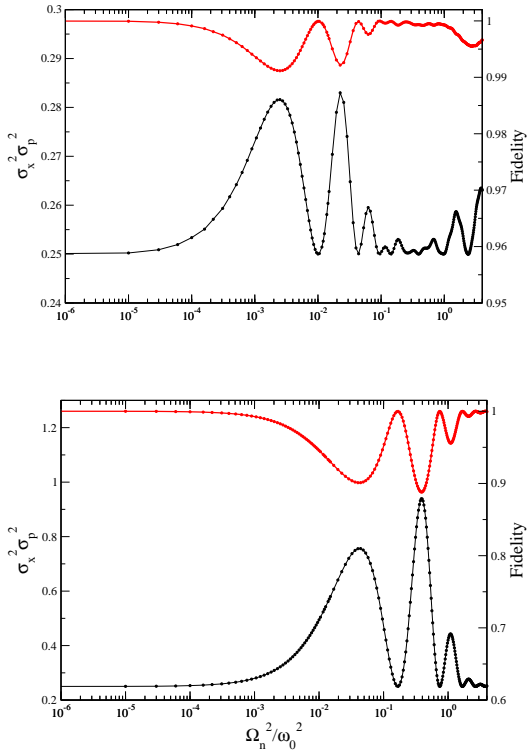

\begin{center}
\includegraphics[width=0.8\columnwidth]{adiabaticnewfig6a.eps}

\vspace{1cm}

\includegraphics[width=0.8\columnwidth]{adiabaticnewfig6b.eps}
\end{center}
\caption{(Color online) The variance product,  $\sigma^2_{x}(t_{\mathrm f}) \sigma^2_{p}(t_{\mathrm f})$ 
 (bottom solid line in black , left vertical axis) and the fidelity (top solid line in red , right vertical axis) as 
a function of the scaled modulation frequency $\Omega_n^2/\omega_0^2$ for fixed modulation amplitude 
$\beta = 0.1$  with  $t_{\mathrm f}$=25 ms (top plot) and $t_{\mathrm f}$=6 ms (bottom plot).}
\label{adiabaticfig6}
\end{figure}
The case of strategies in which there are also time intervals with negative square frequencies is 
more involved, and a typical example is shown in the bottom plot of Fig. 4 for the $t_{\mathrm f}$=6 ms 
case. Multiple peaks now occur -- one located at the minimum negative square frequency, as the system 
spends a large amount of time in that region, and another at a positive square frequency where the plot 
shows a local maximum in the final stage of cooling. The peak at zero frequency gets contributions from 
three distinct time intervals, two with a zero-crossing behavior, before and after reaching 
the minimum square frequency, and one from positive square frequencies alone at the very end 
of the cooling (see Fig. 1 in \cite{Choi}), resulting in a discontinuity in the temporal density of states.   
The high peaks in the $\omega^2 < 0$ region for the $t_{\mathrm f}=$6 ms case indicate that the contribution 
from the antitrapping stage is much more significant than what a simple inspection of the Ermakov 
trajectory would suggest. 
 
The effect of sinusoidal noise was studied by modifying the Ermakov trajectory $\omega^2_{E}(t)$ 
 to ${\tilde{\omega^2}_{E}(t)}=\omega^2_{E}(t) [1 + \beta  \sin(\Omega_{n} t)]$ and evolving the 
Schr\"{o}dinger Equation under this  time-dependent potential. In Fig. \ref{adiabaticfig5} we 
plot the  final variance product, $\sigma^2_{x}(t_{\mathrm f} )  \sigma^2_{p}(t_{\mathrm f} ) $ and the fidelity (inset)
for various $t_{\mathrm f}$ and $\beta$. We have chosen two possible cases of the modulation frequency, 
$\Omega_n < \omega_0$ ($\Omega_{n}  = \omega_0/2$) and  $\Omega_n > \omega_0$   
($\Omega_{n}  = 3\omega_0/2$). The results show that, as to be expected,  
the  deviation from the desired state increases with increasing $\beta$, although 
the effect is now much more pronounced than those for the random noise.   
Again, as we expected from  Fig. \ref{adiabaticfig2},  it is seen that the short $t_{\mathrm f}$ 
cases that involve negative square frequencies are more sensitive to noise.
There is also an overall oscillatory behavior as a function of $t_{\mathrm f}$ for the variance product, and 
this may be explained as follows. If one assumes, as a very crude, first order 
approximation, that the only relevant quantity in the wave function  evolution  is 
the instantaneous trapping frequency in the presence of noise, {\it i.e.} we completely 
disregard the trajectory history,  one can write using Eq. (\ref{varprod}) the final position variance
\begin{equation}
\sigma^2_{x}(t_{\mathrm f}) \sigma^2_{p}(t_{\mathrm f})  = \frac{\hbar^2}{4} 
\frac{\tilde{\omega}_{\mathrm f}}{\omega_{\mathrm f}},
\end{equation}
where $\tilde{\omega}_{\mathrm f}$ is the new final trap frequency  in the presence of the added noise.
Given the form of the noise, we have $\tilde{\omega}_{\mathrm f} = \omega_{\mathrm f} 
\sqrt{ 1 + \beta  \sin(\Omega_{n} t_{\mathrm f}) }$, and  hence
\begin{equation}
\sigma^2_{x}(t_{\mathrm f}) \sigma^2_{p}(t_{\mathrm f})  \sim  \frac{\hbar^2}{4}  
\sqrt{ 1 + \beta  \sin(\Omega_{n} t_{\mathrm f}) }, 
\label{rough}
\end{equation}
so that there is an oscillatory behavior  expected for the variance product as a function of 
$t_{\mathrm f}$ at constant $\Omega_n$ (and vice versa).  This result indicates a fairly strong dependence 
of robustness on $t_{\mathrm f}$ or, to be more precise, on the deviation from the 
desired final trap frequency  due to the presence of noise which, in this case, is given by 
the multiplicative  factor $\sqrt{ 1 + \beta  \sin(\Omega_{n} t_{\mathrm f}) } $.

\begin{figure}[t]
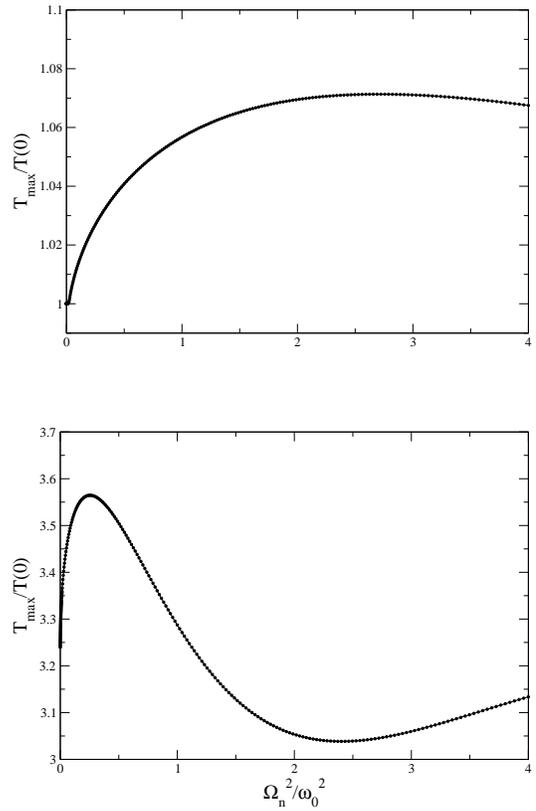

\begin{center}
\includegraphics[width=0.8\columnwidth]{adiabaticnewfig7a.eps}

\vspace{1cm}

\includegraphics[width=0.8\columnwidth]{adiabaticnewfig7b.eps}
\end{center}
\caption{Scaled maximum temperature as a function of the  the scaled modulation frequency 
$(\Omega_n^2/\omega_0^2)$  for fixed modulation amplitude $\beta = 10\%$  with  
$t_{\mathrm f}$=25 ms (top) and $t_{\mathrm f}$=6 ms (bottom).}
\label{adiabaticfig7}
\end{figure}

Next, we have fixed the amplitude of the sinusoidal noise $\beta$ to be 0.1 i.e. 10\% of the 
instantaneous Ermakov trajectory $\omega^2_{E}(t)$, and calculated the response of the system 
to noise in a range of modulation frequencies, $\Omega_{n} \in [0, 2 \omega_0]$. 
This is presented in Fig. \ref{adiabaticfig6} 
for both the $t_{\mathrm f}$=25 ms  and $t_{\mathrm f}$=6 ms cases. 
It is observed that the system is fairly sensitive to the variation of the modulation frequency, and 
it is also evident that the $t_{\mathrm f}$= 6 ms case with the negative square frequency stage
shows a very strong deviation from the  desired state.
As before, this is likely due to the enhanced  vulnerability to 
external modulation during the anti-trapping stage. The oscillatory behavior as a function of the modulation 
frequency $\Omega_n$ for fixed $t_{\mathrm f}$ may, again, be partially  explained using Eq. (\ref{rough}).  
From the same equation, the maximum possible deviation from the  desired state
is expected to be of the order $\beta/2$.  Since we use $\beta $= 0.1 for the maximum noise, 
this rough estimate holds for the $t_{\mathrm f}$=25 ms case but not for the $t_{\mathrm f}$=6 ms case. 
The fact that the maximum possible deviation  in variance product worked 
out from Eq. (\ref{rough}) is smaller than that observed indicates the presence of other contributing 
factors that disrupts the  shortcut to adiabaticity as one would expect. 

In order to produce a strong response,  the 
modulation frequency $\Omega_n$ should, in general, match those  trap frequencies along the trajectory 
that varies slowly {\it i.e.} those with large $dt/d(\omega^2)$ in Fig. \ref{adiabaticfig4}. 
For $t_{\mathrm f}$ = 25 ms this seems to be the case, in that the strongest response 
appears near $\Omega_n \approx 0$. However for $t_{\mathrm f}$ = 6 ms, although a strong response 
is also visible near the expected region of $\Omega_n \approx 0$,  the maximum peak clearly appears near 
$\Omega_n^2/\omega_0^2 \approx 0.5$.  On close examination,  it was found that the 
peaks appear near certain multiples of a characteristic angular frequency 
$\omega_{C} = \pi/ 2 t_{\mathrm f}$, obtainable from Eq. (\ref{rough})  as the first local maximum.  
It was observed that the the first three peaks are found at $\tilde{\Omega} = \omega_C$,  $4\omega_C$, and
 $7\omega_C$.  Comparing the square of characteristic angular frequency, the  $t_{\mathrm f}$ = 25 ms 
case gives $\tilde{\Omega}^2/\omega_0^2 = 0.0016, 0.0256, 0.0784$, while the $t_{\mathrm f}$ = 6 ms 
case gives $\tilde{\Omega}^2/\omega_0^2 =0.0278, 0.4444, 1.3611$, approximately coinciding with 
the corresponding locations of the maximum peaks. This shows that, in addition to the 
temporal density of states, the characteristic angular frequency that depends on the precise 
type of noise and its effect on the final trapping frequency is another important parameter 
in characterizing the robustness of this system.

 Finally, we note that the overall behavior of the fidelity mirrors that of the variance product, indicating that, in our case, 
the amount of squeezing is indeed an efficient alternative parametrization of robustness.

\subsection{Reduction of the maximum temperature}

A crucial finding in our previous work \cite{Choi} was the limitation imposed by the 
high temperature attained by the atoms during the cooling process. Therefore, related to the 
study of robustness of the system to noise,  we have investigated the maximum temperature attained 
by the atoms during the cooling process in the presence of sinusoidal modulation of various modulation 
frequencies. For the Ermakov trajectory which does not contain an antitrapping stage, the sinusoidal 
modulation can provide momentum kicks, increasing the momentum variance.  On the other hand, for 
the $t_{\mathrm f}$=6 ms case that includes an antitrapping stage,  with a modulation of the curvature 
of the anti-trap  one may be able to ``catch" the wave function spread in such a way that the momentum 
variance is reduced. Qualitatively, this may be viewed similarly to the mechanism behind the dynamic 
equilibrium attained by an inverted pendulum with an oscillating pivot. 
Figure \ref{adiabaticfig7} shows the scaled maximum temperature as a function of the modulation frequency. 
It is found that for the $t_{\mathrm f}$=25 ms  case, there is some heating as the modulation 
frequency is increased.  However, interestingly, for the $t_{\mathrm f}=$6 ms case 
that contains the negative square frequency stage, it shows 
a dip in the maximum temperature around the modulation frequency of $\Omega_n =  1.55\omega_0$ to a 
value less than that corresponding to no  sinusoidal modulation ($\Omega_n = 0$). 
In addition, the variance product at this frequency is also very close to the minimum uncertainty 
state, indicating this is the optimum  set-up for the $t_{\mathrm f}$= 6 ms case if reducing atom 
losses is the most important priority. 
This provides a motivation for intentionally adding a sinusoidal modulation in the cooling scheme 
to minimize atom-loss due to heating, adding a further dimension to
the issue of cooling optimization when an antitrapping stage is
present \cite{Stefanatos}. It has been shown \cite{Stefanatos1} that the cooling time is
minimized, in the presence of a trapping stage, using a ``bang-bang''
control, {\it i.e.} with stepwise changes of the trapping frequency,
but the possible addition of a sinusoidal driving may call for a
modification, or at least a check, of this optimization procedure. 
  
\section{Discussion}

The results reported in this paper are encouraging for the experimental 
implementation of  frictionless cooling, possibly including a
negative square frequency stage,  in Fermi-Bose mixtures using optical
dipole traps and optical lattices. A scheme using a single-frequency 
optical dipole trap with a continuously tunable laser is not feasible due to the 
large atom losses expected in crossing the dominant atomic transition 
from the red-detuned to the blue-detuned side to achieve antitrapping.
Nevertheless, this issue may be circumvented through the use of two
laser beams at constant frequencies,  opposite detunings, and
variable power ratio, such as the bichromatic optical dipole traps  proposed 
in \cite{Onofrio} to allow optimal heat capacity matching between the 
Bose and Fermi gases \cite{Presilla,JSP}.
The presence of noise related to power fluctuations and beam-pointing
stability for both laser beams in this configuration makes the
discussion presented in this paper quite relevant for implementing
 frictionless cooling in bichromatic optical dipole traps.
Bichromatic trapping has been recently implemented at the
magneto-optical trap stage for a single species \cite{Qiang}, and two-species 
selective trapping and cooling has been demonstrated with hybrid traps 
involving magnetic and optical confinement \cite{Catani,Baumer}. 
Therefore experiments involving trapping and cooling of two species in 
a bichromatic trap should be within reach in the near future.
Our discussion should  also be relevant to the case of  frictionless
cooling in optical lattices, as recently discussed \cite{Yuce1}.
Dynamically variable spacing (the so-called accordion
lattice) allows for a continuous increase of lattice periodicity, as 
experimentally demonstrated in \cite{Fallani,Li,Assam}, without 
the need to change the laser frequency. In this case an additional
source of noise during the  trap expansion is due to the 
presence of acousto-optic deflectors, adding up to the beam-pointing 
stability of the lasers. Implementing   frictionless cooling with 
a negative square frequency stage via a bichromatic optical lattice 
could also be of great relevance to investigate fundamental issues of 
statistical mechanics as the approach to equilibrium in atomic systems 
at negative temperatures \cite{Mosk,Rapp,Iubini}.

In summary, we have found that the robustness of the  frictionless cooling method to noise
can be analyzed by characterizing the final quantum state in terms of the amount of squeezing  as well as the usual fidelity. 
It was found that the method is quite robust to the presence of broadband noise in the trapping 
frequency, and further analysis involving monochromatic sinusoidal modulation has allowed 
us to resolve its response.  We have studied the dependence of the squeezing  and fidelity on 
the deviation from the expected Ermakov trajectory, and identified the 
role of such measures as the temporal density of states and characteristic angular frequency. 
Numerical simulations indicate that, despite the perception that short $t_{\mathrm f}$ could mean 
less time for noise to interfere with the system, too short a $t_{\mathrm f}$ is best avoided in practice. 
In addition, a way  to reduce the maximum temperature and hence atom losses was found by 
adding a high frequency sinusoidal modulation, which helps to mitigate one of the 
limitations of this scheme found in our previous work.

\end{document}